# α-cluster structure of $^{18}$Ne

M. Barbui[a], A. Volya[a,e], E. Aboud[a,1], S. Ahn[a,2], J. Bishop[a], V.Z. Goldberg[a], J. Hooker[a], C.H. Hunt[a,3], H. Jayatissa[a,4], Tz. Kokalova[d], E. Koshchiy[a], S. Pirrie[d], E. Pollacco[b], B.T. Roeder[a], A. Saastamoinen[a], S. Upadhyayula[a,5], C. Wheldon[d] and G.V. Rogachev[a,c]

[a] *Cyclotron Institute, Texas A&M University, MS3366 College Station, TX, 77843, USA*
[b] *IRFU, CEA, Saclay, Gif-sur-Yvette, France*
[c] *Department of Physics &Astronomy, Texas A&M University, College Station, TX 77843, USA*
[d] *School of Physics and Astronomy, University of Birmingham, Birmingham, United Kingdom*
[e] *Department of Physics, Florida State University, Tallahassee, FL 32306-4350, USA*

**Abstract:** In this work, we study α-clustering in $^{18}$Ne and compare it with what is known about clustering in the mirror nucleus $^{18}$O.
The excitation function for $^{14}$O+α resonant elastic scattering was measured in inverse kinematics using the active target detector system TexAT. The data cover the excitation energy range from 8 to 17 MeV. The analysis was performed using a multi-channel R-matrix approach. Detailed spectroscopic information is obtained from the R-matrix analysis: excitation energy of the states, spin and parity as well as partial alpha and total widths. This information is compared with theoretical models and previous data. Correspondence between the levels in $^{18}$O and $^{18}$Ne is established. We carried out an extensive shell-model analysis of the $^{18}$O and $^{18}$Ne mirror pair. The agreement between theory and experiment is very good and especially useful when it comes to understanding the clustering strength distribution. The comparison of the experimental data with theory shows that certain states, especially at high excitation energies, are significantly more clustered than predicted. This indicates that the structure of these states is collective and is aligned towards the corresponding alpha reaction channel.

[1]*Present address: Nuclear Criticality Safety Division, Lawrence Livermore National Laboratory, Livermore, CA 94550, USA*
[2]*Present address: Institute for Basic Science, Center for exotic nuclear studies, Yuseong-gu, Daejeon, 34126 Korea*
[3]*Present address: FRIB, Michigan State University, East Lansing MI, 48824, USA*
[4]*Present address: Argonne National Laboratory, Lemont, IL 60439, USA*
[5]*Present address: TRIUMF, Vancouver, BC V6T 2A3, Canada*

## Introduction:

Charge independence is an important feature of the nuclear force. This property is broadly used to predict mirror nuclei's structure and evaluate mirror nuclear reaction cross-sections. These predictions are essential and often used in nuclear astrophysics when the direct measurement of one of the nuclear species in the mirror pair is difficult to achieve. At the same time, there are well-known cases of isospin-symmetry breaking; see, for example, refs. [1-3].

Rare isotope beam facilities made it possible to measure nuclear reactions with low-intensity radioactive beams. Resonant scattering reactions studied with thick active or passive targets in inverse kinematics [4] have become a productive tool at these facilities due to their large cross-sections and the possibility to investigate excitation functions in a wide energy range with single beam energy. Proton-rich nuclei have been studied mainly through resonant scattering on hydrogen to investigate low-lying single-particle states. Alpha-cluster states in proton-rich nuclei are much less studied. Strongly clustered states are well known in N=Z nuclei; however, adding extra nucleons makes the many-body states significantly more complicated.

In this work, we study the structure of $^{18}$Ne through the resonant elastic scattering of $^{14}$O on $^{4}$He. The structure of the mirror nucleus $^{18}$O has been studied by several groups [5-14]. In particular, the mirror reaction $^{14}$C+$\alpha$ was previously measured by Avila et al. [5] with high statistics using a high-intensity $^{14}$C beam from the tandem accelerator at FSU John D. Fox laboratory. The detailed R-matrix analysis of the $^{14}$C+$\alpha$ data in ref. [5] provides a perfect starting point for the analysis of the new $^{14}$O+$\alpha$ data. In order to minimize possible systematic effects, the R-matrix analysis of the $^{18}$Ne data is performed with the same R-matrix code used in ref. [5]. Even though large body of experimental data exists for $^{18}$O and some data exist for $^{18}$Ne, especially at the low energies relevant for astrophysics [15-23], a systematic comparison of the two systems in a wide energy range has not yet been made. Fu et al. in ref. [17] attempted a comparison of the alpha cluster states in $^{18}$Ne and $^{18}$O. The authors claimed significant differences between the two mirror nuclei, but the analysis was inconclusive given the limited energy range.

We also compare the experimental data with the predictions of the shell-model calculations performed with a new FSU Hamiltonian [24]. In principle, with a large enough valence space, including the reaction continuum, these calculations should be able to reproduce the alpha-cluster structure of $^{18}$Ne and $^{18}$O and help understand the role of nucleon degrees of freedom in the alpha-cluster structure of N≠Z nuclei. Comparing the results from the R-matrix analysis and the shell-model calculation for each spin parity, we better understand the origin and systematics of the clustering strength.

Other approaches, such as antisymmetrized molecular dynamic calculations (AMD) or the Orthogonality condition model (OCM), can also be used to probe the cluster structure in N≠Z mirror pairs; see for example refs. [25, 26].

## Experimental Setup:

The reaction $^{14}$O + $^{4}$He was studied in inverse kinematics using the TexAT active target [27]. The $^{14}$O beam was produced with the reaction $^{14}$N(p, n)$^{14}$O using Magnetic Achromat Recoil Separator (MARS) at the Cyclotron Institute at Texas A&M University [28]. The $^{14}$N primary beam with energy of 11 MeV/nucleon was delivered by the K500 Cyclotron and directed into a 9.2 cm long liquid nitrogen-cooled gas cell. The pressure of the $H_2$ gas inside the cell was 2280 Torr. The energy of the $^{14}$O beam was 61.8 MeV with an intensity of about $10^4$ pps. The main beam contaminant was $^{7}$Be at the few percent level. A total of $3.8 \cdot 10^8$ $^{14}$O beam ions were delivered to the active target during the experiment.

The Texas Active Target, TexAT, is a gas-filled Time Projection Chamber. The high level of segmentation of the anode combined with the measurement of the electron drift time in the vertical direction provide a 3D reconstruction of the tracks produced by the incoming beam ions and the charged reaction products. Measuring the specific energy loss of the particles in each segment also provides particle identification capability. A complete description of the TexAT device can be found in ref. [2]. In the present experiment, TexAT was operated with a $^4$He (96%) and $CO_2$ (4%) gas mixture. The gas pressure of 580 Torr was enough to stop the $^{14}$O beam before the last $1/8^{th}$ of the active volume. This last portion of the active volume was used as a ΔE detector to identify the alpha-particles hitting the silicon wall. Nine 5x5 $cm^2$ quadrant detectors are arranged in a silicon wall of a thickness (600-700 μm), as shown in FIG. 1. Given the energy of the $^{14}$O beam, the alpha-particles from elastic scattering are stopped in the silicon detectors.

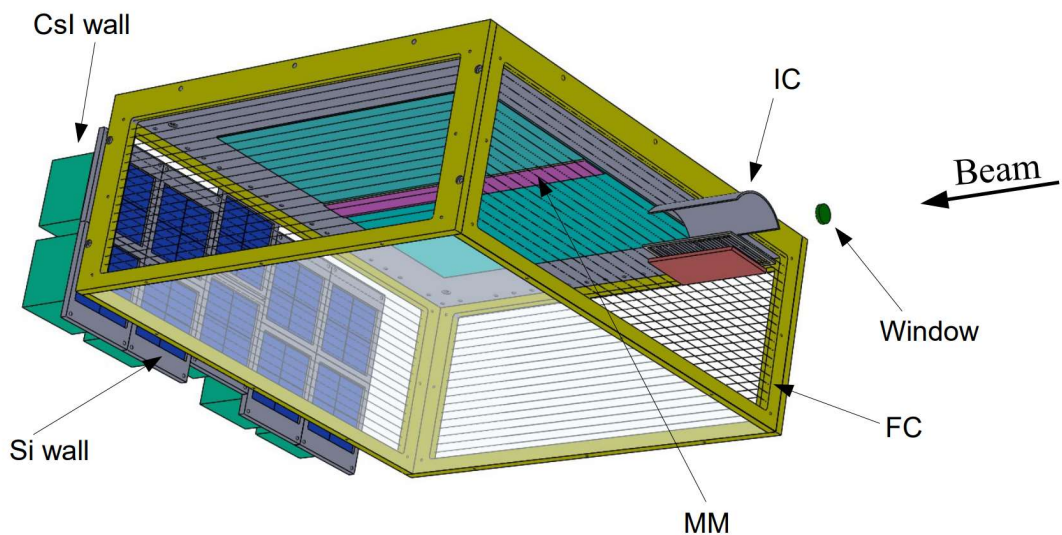

FIG. 1. Drawing of the TexAT assembly (colors online only). The bottom plate has been made transparent to see the interior. The top part is the Micromegas (MM), where the magenta color marks the central pads and the turquoise the side regions. The Si detectors (blue) can be backed by CsI detectors (green) which were not used in this experiment. A field cage (FC) maintains a uniform electric field inside the active volume. The beam travels from right to left along the central pads. An ionization chamber (IC) before the Micromegas is used to count the beam ions.

**Data analysis:**
To implement particle identification, ΔE-E plots were produced using the energy deposited in the last few centimeters of the Micromegas as ΔE and the energy deposited in the Silicon detector as E. As shown in FIG. 2, the alpha-particles can be easily separated from the protons. Alpha-particles were identified with a two-dimensional gate around the alpha-particle locus.

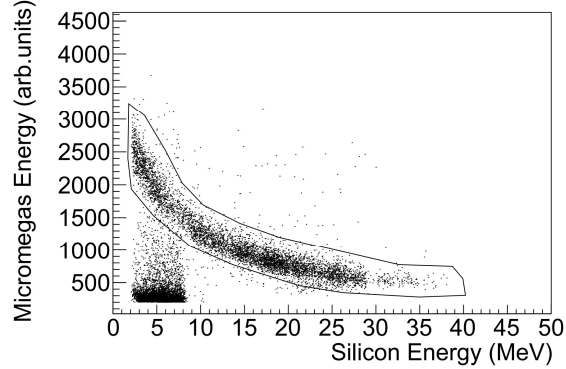

FIG. 2. ΔE vs. residual energy plot for the quadrants positioned at 9° from the entrance window on the left side of the incoming beam. The alpha-particles are selected with the 2-dimensional gate shown by the black line. The protons are in the bottom left corner of the figure.

The tracks of the alpha-particles detected in each silicon detector were reconstructed. The central region of the Micromegas is divided into 6 x 128 pixels. The track reconstruction in this region is done by finding adjacent pixels that fired. The side regions are divided into strips, perpendicular to the beam directions, and chains of pixels (referred to as just chains) parallel to the beam direction. The track reconstruction's first step is matching strips and chains in the side regions of the Micromegas detector. This is done for each event using the timing information. For each strip signal, the corresponding chain should have a signal simultaneously. After this matching, the Hough transformation [29] is used to further clean the tracks from spurious hits produced by random pads firing in coincidence with the track. This procedure is commonly used to optimize the tracking on Time Projection Chambers and is described in detail in ref. [27]. Event by event, the position of the interaction point in the active target is obtained by intersecting the track corresponding to the detected alpha-particle with the track of the beam if the interaction point is in the region covered by the Micromegas or with the ideal beamline if the interaction point was before the beginning of the Micromegas. The position of the reconstructed interaction point as a function of the alpha-particle energy is shown in FIG. 3(a). The vertex position for alpha-particle elastic scattering, calculated from the reaction kinematics at 10 degrees from the center of the entrance window, is also shown in the figure. The events corresponding to the elastic scattering of $^{14}$O on $^{4}$He were selected with the 2-dimensional gate shown in FIG. 3(a). Events corresponding to inelastic scattering are located below this region. For events corresponding to alpha-particles scattered at or near 180° in the center-of-mass and traveling through the central region of the Micromegas, the vertex reconstruction is obtained from the position of the Bragg peak of the recoiling $^{14}$O. For these events, both alpha-particle and $^{14}$O recoil travel at small angles with respect to the beam axis. Using the reaction kinematics and the known energy loss of the particles in the gas, it is possible to correlate the vertex position with the Bragg peak position. Even in this case, the events corresponding to elastic scattering are separable from the inelastic contribution, as shown in FIG. 3(b).

After selecting the alpha elastic scattering events, for each group of quadrant detectors (defined by their angular position from the entrance window), we plotted the reconstructed scattering angle in the laboratory frame of reference as a function of the energy of the detected alpha-particle. An example of these 2-dimensional plots is shown in FIG. 4. The excitation functions are constructed starting from these 2-dimensional plots. The excitation energy of $^{18}$Ne is derived for each energy bin taking into account energy loss along the track and the scattering angle, which is given by the centroid of the projection of the energy bin on the scattering angle axis. The scattering angle uncertainty is given by the standard deviation of the projection on the angle axis. It varies, ranging from 3° at the lowest energy to about 1° at the highest energy. The cross-section is calculated for each energy bin considering the proper scattering angle, the solid angle covered by the detectors, and the number of beam and target particles.

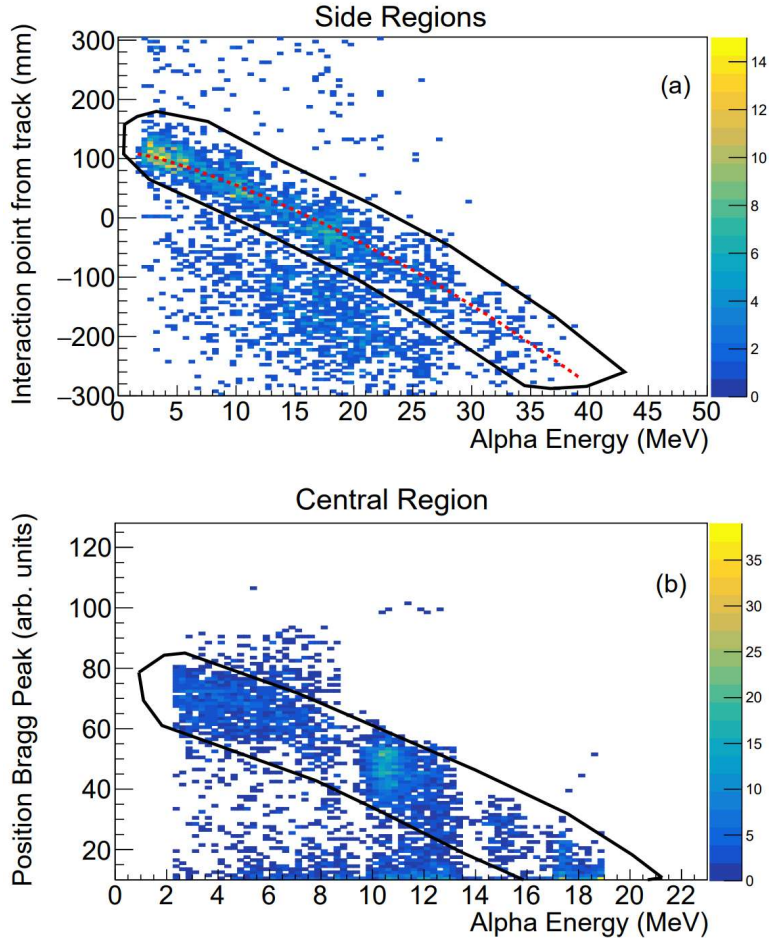

FIG. 3. Vertex reconstruction for alpha-particles detected in the TexAT detector. (a) Alpha-particles traveling in the side regions of the Micromegas; the red dashed line shows the interaction point calculated for elastic scattering in a detector placed at 10 degrees from the beam axis from the center of the entrance window. Zero mm is the beginning of the active volume. The elastic events were selected with a two-dimensional cut (black solid line). The events below the cut are inelastically scattered alpha-particles. (b) Position of the Bragg peak in the Micromegas versus the energy of the alpha-particles traveling in the central region of the Micromegas. Elastic events are selected with a two-dimensional cut (black line). The plot extends up to alpha energies of about 20 MeV; above this energy, the Bragg peak occurs before the beginning of the detector's active volume, and it is not detected. Most inelastic events occur at the beginning of the active region or before it.

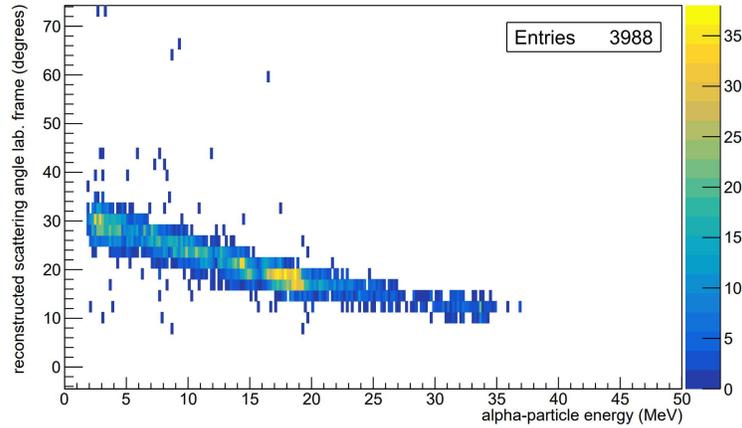

FIG.4. Reconstructed scattering angle as a function of the detected alpha particle energy for the quadrants positioned at 12° from the beam axis at the center of the entrance window.

The alpha elastic scattering excitation functions of $^{18}$Ne are shown in FIG. 5. The excitation function measured by Fu et al. [17] with the Thick Target Inverse Kinematics Technique near zero degrees in the laboratory frame of reference is shown in FIG. 5(b).

A multi-channel R-matrix analysis was performed on the data using the code MinRmatrix [30]. This code is based on the R-matrix theory formulated by Lane and Thomas [31]. Data from ref. [17], FIG. 5(b) were also included in the fit. For this excitation function, the scattering angle for each energy bin in the spectrum was calculated using kinematics and assuming the detector was placed at 1.25° from the beam axis at the entrance window. An error of 30% was assigned to these data to consider uncertainties in the angle reconstruction and avoid an excessive weight of this dataset compared to the present work. The alpha and the proton emission channels were considered in the R-matrix fit. In particular, the alpha decay channel going to the ground state of $^{14}$O and the proton decay channels going to the $^{17}$F ground state (5/2$^+$) and the $^{17}$F first excited state (at 495 keV with 1/2$^+$) were included in the R-matrix analysis. The experimental excitation functions corresponding to $^{14}$O-$^4$He elastic scattering events were used in the R-matrix analysis. The track reconstruction allows us to remove the inelastic events in a model-independent way. Since we do not have proton-emission data, the proton channels included in the fit account for all possible non-elastic scattering cross-sections. Avila et al. [5] reported 54 levels in $^{18}$O in the excitation energy range from 8 to 15 MeV. The initial parameters for our R-matrix analysis (energy eigenvalues and reduced partial width amplitudes) were derived from those used in ref. [5] for $^{18}$O. The spin and parity of each level were assigned as in ref. [5]. Given the low statistics of our dataset, the present experiment is primarily sensitive to states with significant alpha partial width and relatively small proton partial width. Therefore, we initially considered the 37 states listed in ref. [5] with the clear spin-parity assignment and dimensionless alpha reduced width in $^{18}$O larger than 0.02. These initial parameters were fitted to the experimental data without specified limits. The fit was performed with the MINUIT [32] code using the fitting procedure MIGRAD. The fit pushed two 2$^+$ states initially at excitation energies around 13 MeV up to 16.26(2) and 16.9(2) MeV. These levels are discussed in the J$^\pi$ = 2$^+$ paragraph. One of the 4$^+$ states initially located between 10 and 11.5 MeV was pushed down to $8.16^{+0.05}_{-1.1}$ MeV. This state is discussed in the J$^\pi$ = 4$^+$ paragraph. After the fit we further removed 9 states. These were very narrow states, their removal would change the reduced $\chi^2$ by less than 1%. In order to improve the fit in the energy region around 11 MeV we tried to introduce a new level. We manually introduced the level and changed the spin-parity assignment to search for the best match. A 6$^+$ level was the best option to reproduce the cross-section, therefore, we added a $6^+$ state at 11.2 MeV and used the fit to obtain the best parameters. This state is discussed in the $J^\pi = 6^+$ paragraph.

The final fit had a total of 29 levels with a reduced $\chi^2$ of 2.7. The results from the fit are reported in TABLE I and compared with the levels in $^{18}$O from ref. [5]. Each state's significance level (S.L.) is reported in the last column of TABLE I. To calculate these values, we used the ratio of the $\chi^2$ obtained from the refit of the data after removing the state and the original (best fit) $\chi^2$. We then compared these ratios with the F-distribution.

To make the comparison between $^{18}$Ne and $^{18}$O easier, the dimensionless reduced width for the α channel, $\theta^2_\alpha$, is calculated for each level as $\theta^2_\alpha = \gamma^2_\alpha/\gamma^2_{SP}$, where $\gamma^2_\alpha$ is the α reduced partial width, and $\gamma^2_{SP} = \hbar^2/\mu R^2$ is the single-particle limit [31] calculated at channel radius R=5.2 fm. The same channel radius was used in ref. [5] for $^{18}$O. Levels with a pronounced alpha-cluster structure, characterized by a large dimensionless reduced width, are highlighted in bold in TABLE I. As two proton channels are considered in the R-matrix fit, the dimensionless reduced proton width, $\theta_p^2$, shown in TABLE I, is calculated as $\theta^2_p = (\gamma^2_{p_0} + \gamma^2_{p_1})/\gamma^2_{SP}$, where $\gamma^2_{p_0}$ and $\gamma^2_{p_1}$ are proton reduced partial widths for the channels populating the ground and the first excited states of $^{17}$F respectively.

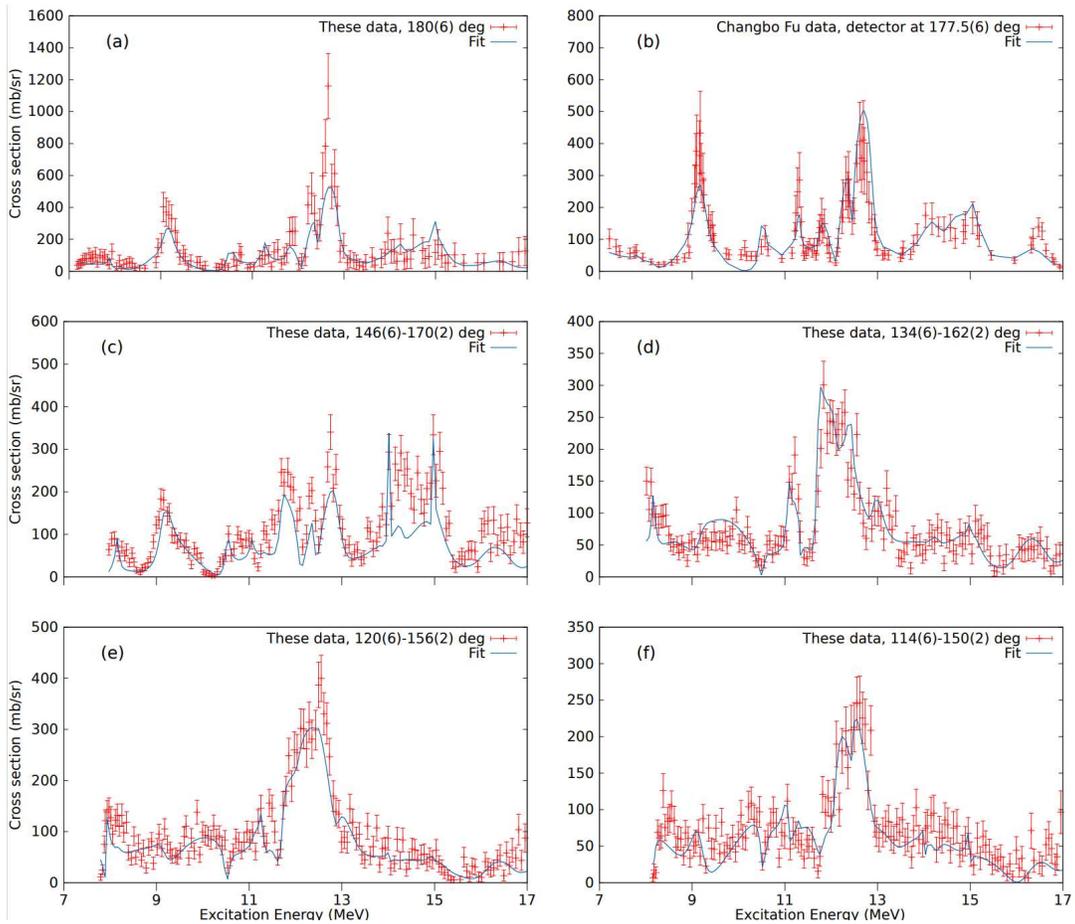

FIG. 5. Excitation function of $^{18}$Ne measured at different angles in the center-of-mass framework. Experimental data are marked in red. Data from ref. [17] are shown in panel (b), whereas the other panels show only data from the present work. The solid blue line shows the R-matrix best fit.

**Results and comparison with the shell-model calculation:**

It is clear from TABLE I that there is a good correspondence between the levels observed in $^{18}$O and those in $^{18}$Ne, especially for the levels with significant alpha spectroscopic factors in the mirror pair.
To investigate the properties of the measured excited states in more detail, the experimental results were compared with the shell-model calculation [24]. This model uses the harmonic oscillator (HO) basis to treat cluster configurations. In this way, we can retain complete control of the center-of-mass and the relative motion between clusters, preserving the Pauli antisymmetry concerning all nucleons. Another practical convenience comes from the algebraic properties of the harmonic oscillator [33] that allow using the number of excitation quanta to characterize each cluster and the relative motion between clusters.
While many-body configurations are generally mixed, a recent study of $^{19}$F [34], performed with the same theoretical approach, shows that the oscillator quantum numbers are still useful in understanding the nature of the clustered states and their properties. Thus, the strategy of our theoretical approach is to build a particle-hole excitation hierarchy using the oscillator quantum numbers, accounting for the alpha coupling strength. We use the FSU interaction Hamiltonian [35, 36] that considers the s, p, sd and fp valence shells. States with the same particle-hole excitation in each $J^\pi$ group have the same number of $\hbar\omega$ excitations in the relative motion HO and the same number of nodes, n, in the radial wave function. The number of nodes in the radial wave function is directly correlated to the number of excitation quanta, Q, that the alpha-particle needs to take away in the decay to reach the $^{14}$O in its ground state (Q = 2n + L, where L is the orbital angular momentum). For example, the $0^+$ ground state and the second $0^+$ excited state in $^{18}$Ne have $0\hbar\omega$ excitations and n = 3 in the model. This means that the alpha-particle in the decay needs to take away 6 excitation quanta Q to reach the $^{14}$O in its ground state. The alpha wave function with n = 3 has the largest overlap with the $^{18}$Ne configuration, where an alpha-particle is made with two nucleons from the sd shell (4 quanta) and two nucleons from the p shell (2 quanta). Particle-hole mixing could be considered as it was done in ref. [37], however, we do not consider it here. The FSU Hamiltonian [35, 36] adopted in this work has been shown to work well without mixing [34]. We emphasize that without mixing each state with a given particle-hole structure (given by the number of $\hbar\omega$ excitations) can only couple to a unique alpha-cluster channel. This restriction may potentially lead to discrepancy between the model calculations and the experimental data for the states with mixed nature. Different alpha decay channels are identified by the number of nodes in the relative alpha plus core wave function. The cumulative decay strength, sum of spectroscopic strength, is normalized by the number of such channels. It can also be expected that the alpha decay of states in a certain energy region would be dominated by a specific channel, see Ref. [24] for theoretical discussion. The results of the shell-model calculation are shown in TABLE II and compared with the experiment. In the following, we discuss the experimental results and the comparison with the shell-model by $J^\pi$ groups.

**$J^\pi = 0^+$**

A wide $0^+$ state is found in $^{18}$Ne at an excitation energy of 9.8(8) MeV, corresponding to a wide $0^+$ state observed in $^{18}$O [5] at 9.9(1) MeV excitation energy. In ref. [8] Johnson et al. showed that this state could be described by a single alpha-particle plus $^{14}$C core. This s-wave scattering feature has smooth phase shift energy dependence only reaching a maximum of 45 degrees. Yet, it is necessary to reproduce the $^{14}$C+α excitation functions [5,8]. A similar feature is also necessary to fit the $^{14}$O+α excitation functions. To verify that the best fit s-wave $^{14}$O+α phase shift has the same nature as for the $^{14}$C+α, we followed the procedure described in [8] and used a simple alpha plus core potential model to calculate the phase shift for this broad L = 0 state. We tuned the depth of the potential to obtain a relative motion wave function with 5 nodes (as indicated by the shell-model). We compared the sine squared of the phase shift obtained in the calculation with the same quantity obtained by the R-matrix fit. Comparing the R-matrix results with the potential model calculation is easy for this state because there are no other $0^+$ states in the energy region, and the proton channel is highly suppressed.

TABLE I: Summary of the resonance parameters observed in the elastic scattering of $^{14}$O on $\alpha$. The resonance parameters obtained from the elastic scattering of $^{14}$C on $\alpha$ [5] are also shown for comparison. The * indicates states from ref. [18]. $E_{exc}$ is the excitation energy of the state, $J^\pi$ indicates spin and parity, $\Gamma_{tot}$ is the total width of the state, $\Gamma_\alpha$, $\Gamma_p$, $\Gamma_n$ are the partial alpha, proton, and neutron widths, respectively, $\theta^2_\alpha$ is the alpha dimensionless reduced width, $\theta^2_p$ is the proton dimensionless reduced width, $\theta^2_n$ is the neutron dimensionless reduced width. States with $\theta^2_\alpha > 0.1$ are highlighted in bold. The last column shows the significance level (S.L.) for every observed state in $^{18}$Ne. The particle thresholds for alpha, proton, and two proton emissions in $^{18}$Ne are 5.112 MeV, 3.992 MeV, and 4.522 MeV, respectively [18]. The particle thresholds for an alpha, neutron, and two neutron emissions in $^{18}$O are 6.227 MeV, 8.044 MeV, and 12.1890 MeV respectively [18].

| $J^\pi$ | This work $^{18}$Ne | | | | | | $^{18}$O | | | | | | S.L. |
|---|---|---|---|---|---|---|---|---|---|---|---|---|---|
| | $E_{exc}$ MeV | $\Gamma_{Tot}$ keV | $\Gamma_\alpha$ keV | $\Gamma_p$ keV | $\theta^2_\alpha$ | $\theta^2_p$ | $E_{exc}$ MeV | $\Gamma_{Tot}$ keV | $\Gamma_\alpha$ keV | $\Gamma_n$ keV | $\theta^2_\alpha$ | $\theta^2_n$ | |
| $0^+$ | 9.8(3) | 4200 | 4200 | 0 | **1.4(6)** | 0 | 9.9(1) | 3200 | 3200 | 0 | **1.9(5)** | 0.00 | 90 |
| $1^-$ | | | | | | | 8.04(2) | 2 | 2 | 0 | 0.02(1) | 0.00 | |
| $1^-$ | 9.13(2) | 990 | 390 | 600 | **0.22(2)** | 0.11 | 9.19(2) | 220 | 200 | 20 | **0.20(1)** | 0.07 | 71 |
| $1^-$ | 9.61(2) | 1640 | 1120 | 520 | **0.52(5)** | 0.08 | 9.76(2) | 700 | 630 | 70 | **0.46(4)** | 0.06 | >99.9 |
| $1^-$ | 10.56(4) | 380 | 320 | 60 | **0.11(5)** | 0.01 | 10.8(3) | 690 | 630 | 60 | **0.29(4)** | 0.02 | 70 |
| $1^-$ | 11.74(5) | 360 | 310 | 50 | 0.09(1) | 0.01 | 11.67(2) | 200 | 120 | 80 | 0.04(1) | 0.04 | 87 |
| $1^-$ | | | | | | | 12.12(1) | 410 | 50 | 360 | 0.020(4) | 0.07 | |
| $1^-$ | | | | | | | 12.5(1) | 900 | 300 | 600 | 0.08(3) | | |
| $1^-$ | | | | | | | 13.33(2) | 300 | 30 | 270 | <0.01 | | |
| $1^-$ | 13.73(1) | 1200 | 780 | 410 | **0.2(1)** | 0.04 | 14.3(3) | 900 | 400 | 500 | **0.10(4)** | | 97 |
| $1^-$ | | | | | | | 14.5(2) | 450 | 230 | 220 | 0.05(2) | 0.04 | |
| $1^-$ | | | | | | | | | | | | | |
| $2^+$ | | | | | | | | | | | | | |
| $2^+$ | (7.93(2)) | 75 | 50 | 25 | 0.12(5) | 0.01) | 8.22(1) | 1.9 | 1.7 | 0.2 | 0.030(2) | 0.01 | 62 |
| $2^+$ | 9.21(3) | 540 | 270 | 270 | **0.21(2)** | 0.04 | 9.79(6) | 170 | 90 | 80 | **0.10(3)** | 0.08 | 98 |
| $2^+$ | | | | | | | 10.42(2) | 180 | 40 | 140 | 0.030(8) | 0.04 | |
| $2^+$ | | | | | | | 10.98(4) | 280 | 20 | 206 | 0.010(5) | 0.14 | |
| $2^+$ | | | | | | | 11.31(8) | 250 | 90 | 160 | 0.020(7) | 0.09 | |
| $2^+$ | 10.8(1) | 1580 | 1350 | 230 | **0.55(3)** | 0.02 | 12.21(8) | 1100 | 1000 | 100 | **0.37(9)** | 0.01 | >99.9 |
| $2^+$ | 13.4(2) | 1800 | 1750 | 50 | **0.45(8)** | <0.01 | 12.8(3) | 4800 | 4800 | 0 | **1.56(13)** | 0.00 | 99 |
| $2^+$ | | | | | | | 12.90(3) | 310 | 285 | 25 | 0.090(9) | 0.01 | |
| $2^+$ | | | | | | | 13.17(3) | 150 | 130 | 20 | 0.04(1) | 0.00 | |
| $2^+$ | | | | | | | 13.38(2) | 250 | 220 | 40 | 0.07(1) | 0.00 | |
| $2^+$ | | | | | | | 13.69(1) | 530 | 40 | 490 | 0.010(5) | 0.12 | |
| $2^+$ | | | | | | | 14.12(7) | 160 | 100 | 60 | 0.030(9) | 0.00 | |
| $2^+$ | 16.26(2) | 1100 | 1000 | 100 | **0.2(1)** | 0.01 | | | | | | | 97 |
| $2^+$ | (16.9(2) | 2400 | 1500 | 900 | **0.3(2)** | 0.07) | | | | | | | 63 |
| $3^-$ | | | | | | | 8.29(6) | 8.5 | 2.9 | 5.6 | **0.18(1)** | 0.01 | |
| $3^-$ | 8.76(8) | 870 | 440 | 430 | **1.0(4)** | 0.03 | 9.35(2) | 180 | 110 | 70 | **0.48(13)** | 0.07 | >99.9 |
| $3^-$ | | | | | | | 9.7(1) | 140 | 15 | 125 | 0.040(5) | 0.24 | |
| $3^-$ | | | | | | | 10.11(1) | 16 | 7 | 9 | 0.010(3) | 0.10 | |
| $3^-$ | | | | | | | 10.4(1) | 70 | 50 | 20 | 0.030(3) | 0.02 | |
| $3^-$ | | | | | | | 11.62(3) | 150 | 30 | 120 | 0.010(2) | 0.01 | |
| $3^-$ | 11.0(1) | 1130 | 380 | 750 | **0.21(7)** | 0.09 | 11.95(1) | 560 | 300 | 260 | **0.17(2)** | 0.12 | >99.9 |
| $3^-$ | 12.13(4) | 280 | 180 | 100 | 0.07(3) | 0.01 | 12.64(1) | 110 | 10 | 100 | <0.01 | | 85 |
| $3^-$ | 12.45(2) | 390 | 180 | 210 | 0.07(3) | 0.03 | 12.71(2) | 300 | 120 | 180 | 0.050(4) | | 80 |
| $3^-$ | 12.7(2) | 2300 | 2000 | 240 | **0.7(2)** | 0.02 | 12.98(4) | 1040 | 770 | 270 | **0.32(5)** | 0.10 | 97 |
| $3^-$ | | | | | | | 13.96(2) | 150 | 80 | 70 | 0.030(4) | | |
| $3^-$ | 14.8(2) | 5300 | 4000 | 1300 | **1.0(2)** | 0.12 | 14.0(2) | 2600 | 2100 | 500 | **0.7(1)** | | 88 |
| $3^-$ | | | | | | | | | | | | | |
| $4^+$ | $(8.16^{+0.05}_{-1.1}$ | 90 | 30 | 60 | **0.8(3)** | 0.03) | $7.11^*$ | | | | | | >99.9 |
| $4^+$ | | | | | | | 10.29(4) | 29 | 19 | 10 | 0.09(1) | 0.00 | |
| $4^+$ | | | | | | | 11.43(1) | 40 | 30 | 10 | 0.05(2) | 0.00 | |
| $4^+$ | | | | | | | 12.54(1) | 6 | 5 | 1 | <0.01 | 0.02 | |
| $4^+$ | 13.3(3) | 870 | 850 | 20 | **0.37(4)** | <0.01 | 13.46(2) | 540 | 210 | 330 | **0.12(1)** | 0.06 | 79 |
| $4^+$ | | | | | | | 13.89(1) | 24 | 14 | 10 | 0.010(4) | 0.01 | |
| $4^+$ | | | | | | | 14.52(1) | 250 | 80 | 170 | 0.030(4) | 0.02 | |
| $4^+$ | 14.15(21) | 620 | 380 | 250 | **0.14(10)** | 0.03 | 14.77(5) | 680 | 680 | 2 | **0.28(2)** | 0.00 | 71 |
| $5^-$ | 11.31(4) | 65 | 15 | 50 | 0.03(2) | 0.02 | 11.63(1) | 40 | 30 | 10 | **0.13(1)** | 0.02 | 70 |
| $5^-$ | | | | | | | 12.34(1) | 39 | 26 | 13 | 0.060(5) | 0.01 | |
| $5^-$ | | | | | | | 12.94(1) | 40 | 15 | 25 | 0.020(3 | | |
| $5^-$ | 12.9(2) | 670 | 530 | 140 | **0.48(12)** | 0.04 | 13.08(1) | 180 | 120 | 60 | **0.17(1)** | 0.03 | 85 |
| $5^-$ | | | | | | | 13.82(1) | 25 | 3 | 22 | <0.01 | 0.01 | |
| $5^-$ | 13.79(8) | 290 | 220 | 70 | **0.14(10)** | 0.02 | 14.1(1) | 560 | 260 | 300 | **0.23(2)** | 0.14 | 77 |
| $5^-$ | 14.6(7) | 1180 | 520 | 660 | **0.27(20)** | 0.14 | 14.7(1) | 280 | 230 | 50 | **0.16(6)** | 0.02 | 74 |
| $5^-$ | (14.9(1) | 90 | 60 | 30 | 0.03(2) | 0.01) | 14.82(7) | 140 | 100 | 40 | 0.07(3) | 0.01 | 51 |
| $6^+$ | (11.23(8) | 15 | 5 | 10 | 0.04(3) | 0.02) | | | | | | | 66 |
| $6^+$ | 11.8(2) | 260 | 40 | 220 | **0.23(7)** | 0.34 | 11.69(5) | 23 | 12 | 11 | **0.23(1)** | 0.14 | 99 |
| $6^+$ | 12.4(2) | 350 | 170 | 180 | **0.56(26)** | 0.22 | 12.57(1) | 70 | 50 | 20 | **0.38(8)** | 0.16 | 96 |

It is clear from the results shown in FIG. 6 that the simple potential model reproduces the energy and the width of this broad state well, indicating that the description of this scattering feature as an alpha-particle plus $^{14}$O is valid.

A more repulsive Coulomb interaction in $^{18}$Ne pushes the s-wave scattering feature toward higher center-of-mass energies compared to the $^{18}$O. As a result, the s-wave phase shift peaks at lower values, 23°, in $^{18}$Ne. The R-matrix fit has a large uncertainty band (FIG. 6). Yet, this fit indicates that this scattering feature is present in the $^{14}$O+α excitation functions at the 90% significance level.

When comparing the shell-model with the experimental data, it is essential to look at the whole sequence of the states. Experimentally, only the ground state and two $0^+$ excited states at 3.6 and 4.6 MeV are firmly established. The shell-model reproduces these states well and predicts them to be clustered. The ground state and the second excited state at 4.6 MeV share the clustering strength into the L = 0, n = 3 alpha-clustering channel. The normalized spectroscopic factors for this channel are 0.74 and 0.20, respectively. The first excited state at 3.6 MeV (3.4 MeV in the shell-model calculation) is dominated by the two particle-hole excitation ($2\hbar\omega$) and couples to the L = 0, n = 4 alpha channel with a spectroscopic factor of 0.62; according to the shell-model the remaining strength in the n=4 channel is fragmented. The next $0^+$ state is predicted at about 7.8 MeV of excitation. Here the density of the states becomes high. The theory predicts eight $0^+$ states in the excitation energy range between 7.5 and 12.5 MeV with different particle-hole excitation structures. These states could mix and give rise to the experimentally observed state at 9.8 MeV. It is important to note that the theory predicts two four particle-hole excited states in this region, with $4\hbar\omega$. This is almost certainly the reason for the considerable alpha channel coupling strength that goes into the n = 5 channel. Our unmixed calculations predict that most of the alpha coupling strength in the n = 5 channel will be at about 14 MeV of excitation (SF = 0.15), not at 10 MeV, as seen in the experiment. We expect that particle-hole mixing, channel mixing, and inclusion of the continuum effects would resolve this discrepancy. In particular, the superradiance mechanism [38] facilitates the mixing so that a single state absorbs all the decay width into a given channel. Therefore, we believe that the 9.8 MeV state is a superradiant state that decays into the L= 0, n = 5 alpha-particle channel.

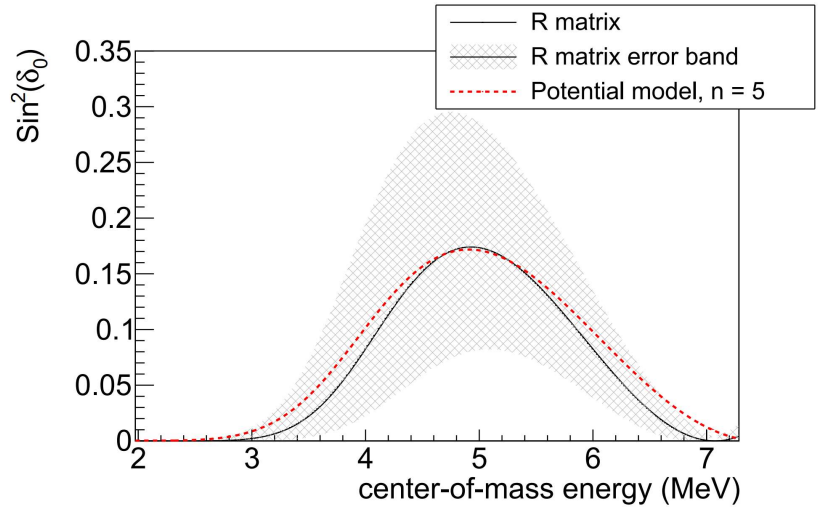

FIG. 6. Comparison of the phase shift squared sine obtained from the R-matrix fit (solid black line) and from a simple potential model calculation with alpha-particle plus core. The dashed red line shows the calculation with n=5 nodes in the radial wave function.

**$J^\pi = 1^-$**

The R-matrix fit of the $^{18}$Ne data shows four $1^-$ levels with pronounced alpha-cluster structure at 9.13(2), 9.61(3), 10.56(4) and 13.73(1) MeV. These levels appear at excitation energies corresponding to the known alpha-cluster states in $^{18}$O. The dimensionless alpha reduced width of these states is compatible within uncertainties with that of the corresponding levels in $^{18}$O, except the state at 10.58 MeV, that has a more considerable alpha strength in $^{18}$O compared to $^{18}$Ne.

The shell-model predicts the five lowest $1^-$ states being $1\hbar\omega$ and thus coupling to the L=1, n=3 alpha channel; four of those states at excitation energies of 4.5, 6.7, 7.4, and 9 MeV should be identified with the ones observed at 4.5, 6.2, 7.7 and 9.13 MeV, respectively. These states saturate all the alpha-clustering strength into the channel with n = 3 nodes. Next, at 9.4, 9.8, and 10.4 MeV, the theory predicts $3\hbar\omega$ states that couple to the n = 4 alpha channel. The spectroscopic factors are large for the 9.4 and 10.4 MeV states, 0.47 and 0.13, respectively. This theory prediction agrees remarkably well with the experimental observation suggesting that 9.61 and 10.56 MeV states share the n = 4 clustering strength.

These low-lying $1^-$ states nearly saturate the alpha decay strength into the n = 3 and n = 4 channels. The next $5\hbar\omega$ states that couple to an alpha channel with n = 5 appear in the shell-model at 11.8 and 13.05 MeV. In the experiment, we observe a level at 13.73 MeV; given its relatively small dimensionless alpha reduced width, this state can couple with the n = 4 or n = 5 channel.

**$J^\pi = 2^+$**

The $J^\pi = 2^+$ levels in $^{18}$Ne are dominated by four broad resonances with a large dimensionless reduced width at 9.21(3), 10.8(1), 13.4(2), and 16.26(2) MeV. A fifth level (16.9(2) MeV) with a pronounced alpha-cluster nature mimics the effect of the continuum at high energy. This latter state has some influence only on the high energy portion of the excitation functions.

The first three broad levels observed in this work have good correspondence with the three broad levels in the $^{18}$O spectrum, although the alpha strength seems to be distributed in a slightly different way in $^{18}$Ne. The level at 16.26 MeV is necessary to reproduce the structure observed in the excitation function between 16 and 17 MeV at various angles. The first $2^+$ level at (7.93) MeV was introduced for symmetry with $^{18}$O. A $2^+$ state at 7.37 MeV, corresponding to the 8.21 MeV state in $^{18}$O, was also reported by Harss et al. [23] in $^{18}$Ne. This state has some influence on the fit, but it is located near the low-energy edge of our excitation functions where the data are incomplete. Therefore, we report it in parentheses in TABLEs I and II.

The shell-model analysis of the $2^+$ states shows some similarities with that of the $0^+$ levels. The first two levels predicted by the shell-model at 2.0 and 4.4 MeV correspond to experimentally established levels at 1.9 and 3.6 MeV in $^{18}$Ne. These levels couple to the alpha channel with n = 2. With spectroscopic factors of 0.64 and 0.15, respectively, these two levels account for most of the alpha strength in the $0\hbar\omega$ group. The third level, predicted at 5.1 MeV with a spectroscopic factor of 0.65, is a $2\hbar\omega$, n=3 state corresponding to the state observed at the same energy in $^{18}$Ne. The shell-model predicts two states with $0\hbar\omega$ and n = 2 at 9.9 and 11 MeV with spectroscopic factors of 0.08 and 0.1, respectively. These two levels most likely correspond to the level measured at 9.21(3) MeV with a dimensionless reduced width of 0.21(2).

In the energy region from 9 to 15 MeV, the shell-model calculation predicts several levels with 2 and $4\hbar\omega$. In this energy range, the calculation fails to predict the magnitude of the spectroscopic factors. The levels measured at 10.8(1) and 13.4(2) MeV have a large dimensionless reduced width (of around 0.5). Given that the levels almost saturate the strength of the $0\hbar\omega$ at lower energy, these two broad states can either: both belong to the $4\hbar\omega$ configuration or one to the $2\hbar\omega$ and the other to the $4\hbar\omega$. The level measured at 16.26 MeV seems to have a good match in the $0\hbar\omega$ state predicted at 16.2 MeV.

**J$^\pi$ = 3$^-$**

A good correspondence is found between the J$^\pi$ = 3$^-$ levels in $^{18}$Ne and $^{18}$O. We observed four levels with relevant alpha-cluster nature at 8.76(8), 11.0(1), 12.7(2) and 14.8(2) MeV. Each of these four levels has a corresponding level in $^{18}$O, with a large dimensionless alpha reduced width. A fifth level corresponding to the 8.29 MeV state seen in $^{18}$O could be added to the fit, but this level is not necessary to reproduce the experimental data. Moreover, this state would be in the energy range where our data are incomplete, and the fit has larger uncertainty. As was the case for the 2$^+$ states, the alpha strength for the 3- states is distributed differently in $^{18}$Ne than in $^{18}$O. In particular, the 8.76 and 12.7 MeV levels in $^{18}$Ne are stronger than the corresponding levels in $^{18}$O.

As shown in TABLE II, the first two levels predicted by the shell-model at 5.1 and 6.4 MeV are in excellent agreement with the two levels known from the literature at 5.1 and 6.3 MeV. The model predicts a prominent alpha-cluster structure for the level at 5.1 MeV with an alpha spectroscopic factor of 0.49. According to the theory, these two levels couple to the alpha channel with n = 2 nodes (1$\hbar\omega$). The model predicts a significant spectroscopic factor (0.33) for the 3$\hbar\omega$ level at 10.7 MeV that corresponds to the alpha-cluster state at 11 MeV ($\theta^2_\alpha$ = 0.21(7)) in this work. The rest of the strength of the one and 3$\hbar\omega$ states in the model is fragmented, while in the experiment, we observe two states with very large alpha-strength at 8.77 and 12.7 MeV. This considerable alpha strength suggests that one of these levels should belong to the 1$\hbar\omega$ configuration and the other to the 3$\hbar\omega$ configuration. Given the good match with the theory, the level at 11.0 MeV should belong to the 3$\hbar\omega$, n = 3 configuration.

The shell-model calculation also predicts the appearance of a 5$\hbar\omega$ level at 12.4 MeV. In the experiment, we see a state at an excitation energy of 14.8 MeV with $\theta^2_\alpha$ = 1 $\pm$ 0.2. Given the high value of $\theta^2_\alpha$, this state could correspond to the 5$\hbar\omega$ level predicted by the model at 12.4 MeV.

**J$^\pi$ = 4$^+$**

In $^{18}$Ne we observed three resonances with a large dimensionless reduced alpha width at (8.16), 13.3(3) and 14.2(2) MeV. The (8.16) MeV 4$^+$ level was not observed in $^{18}$O by Avila et al. [5]. Our fit indicates the need for a 4$^+$ state with large alpha-strength on the low energy side of the spectra and locates the state at an energy of 8.16$^{+0.05}_{-1.1}$ MeV. Due to the high threshold in some detectors, our angular distributions in the energy range from 7 to 8 MeV are incomplete. Therefore, we consider the energy fitted for this level to be an upper limit for the state's position. We believe this state should correspond to the known 7.11 MeV state in $^{18}$O. Considering the error bars, this state could correspond to the 4$^+$ state observed by B. Harss et al. [23] in $^{18}$Ne at 7.05(10) MeV with $\theta^2_\alpha$ = 0.5(2). Given the significant uncertainties on the position and width of this state, we report it in parentheses in TABLEs I and II. The levels at 13.3 and 14.2 MeV have a corresponding level in $^{18}$O with a similar $\theta^2_\alpha$. A level corresponding to the known 10.3 MeV state can be added to the fit, but it is not necessary to reproduce the experimental data.

The shell-model calculation reproduces the energy of the first 4$^+$ excited state, known from the literature at 3.4 MeV. In the calculation, this state has 0$\hbar\omega$ and n = 1. The subsequent 4$^+$ state is predicted at 7.9 MeV with 2$\hbar\omega$ and n = 2. This state corresponds well to the state observed in the present work at 8.16 MeV. The third calculated state is at 9 MeV again with a 0$\hbar\omega$ and n = 1 configuration. We do not observe a level at this energy; neither is it observed in $^{18}$O. The model predicts the first three states to have large spectroscopic factors of 0.43, 0.45, and 0.57, respectively. In our experiment, the level at 8.16 MeV has a large $\theta^2_\alpha$. The two n = 1 states in the calculation account for all the alpha strength in the 0$\hbar\omega$ channel.

Between 10 and 13 MeV, the model predicts several states with $2\hbar\omega$ and n = 2. Above 13 MeV, all states are expected to belong to the $4\hbar\omega$ and n = 3 configuration.

**$J^\pi = 5^-$**

The R-matrix analysis of the 5$^-$ levels in $^{18}$Ne shows three levels with significant alpha-cluster structure at 12.9(2), 13.79(8) and 14.6(7) MeV. These three levels correspond well with levels with large dimensionless reduced widths in $^{18}$O. Although the level at 11.31(4) MeV does not have a significant alpha-cluster structure ($\theta^2_\alpha = 0.03$) in $^{18}$Ne, its presence has some influence on the fit.

Since there are no known 5$^-$ levels in $^{18}$Ne, TABLE II shows the first three 5$^-$ levels observed in the mirror nucleus $^{18}$O. The shell-model reproduces the energy of these first three levels in $^{18}$O quite well, suggesting that they belong to a $1\hbar\omega$ configuration coupling with the n=1 alpha channel. The first two levels have large spectroscopic factors (0.32 and 0.36, respectively).

The fourth level predicted by the model is a $3\hbar\omega$, n = 2 state at 11.8 MeV that matches the level measured at 11.31 MeV. After that, the model predicts a $1\hbar\omega$, n = 1 state that seems the best match for the state observed at 12.9 MeV. From 13 MeV to 15.5 MeV, the model predicts several levels with $1\hbar\omega$ and $3\hbar\omega$. Interestingly, the $3\hbar\omega$ level at 13.7 MeV in the calculation has a relatively large spectroscopic factor (0.25). Using predictions from the model, we suggest that the levels at 13.79 and 14.6 MeV are $3\hbar\omega$, n = 2 states, whereas the weak level at 14.9 MeV can be either a $1\hbar\omega$ or $3\hbar\omega$ level.

**$J^\pi = 6^+$**

Three levels with $J^\pi = 6^+$ are found in $^{18}$Ne. The first level was not seen in ref. [5] and has a small alpha-cluster component. Most likely, the width of the corresponding state in $^{18}$O would have been too narrow to be detected in an experiment performed by Avila et al. [5]. This narrow state is essential to the fit. Without this state, the reduced $\chi^2$ of the fit will change from 2.7 to 3. The effect of removing this state is shown in FIG. 7 for the excitation function at 162°. However, since our statistics are low, more data are required to confirm the existence of this state. We, therefore, present it in parentheses in TABLEs I and II. The levels measured at 11.8(2) and 12.4(2) MeV have significant alpha-cluster structure and corresponding levels in $^{18}$O. The theory here predicts a strongly clustered state (the second 6$^+$ state with a spectroscopic factor of 0.45) at 12.4 MeV with $0\hbar\omega$ and n = 0, and a $2\hbar\omega$, n=1 state at 13.5 MeV with negligible SF. It has been discussed in [5] that configuration mixing may be essential for these two states. The result of this mixing is sharing of alpha-cluster strength between these two states, as observed experimentally.

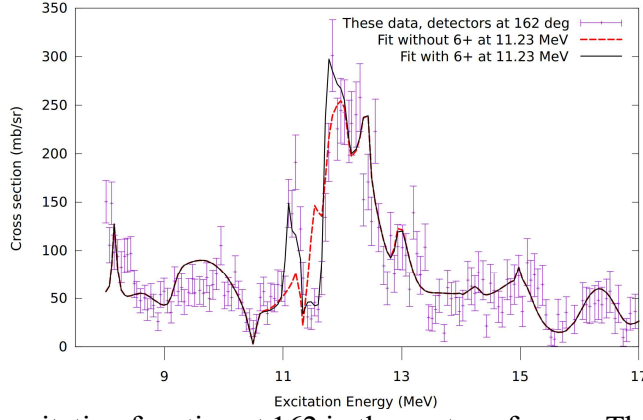

FIG. 7. R-matrix fit of the excitation function at 162 in the center-of-mass. The points are the experimental data, the continuous black line is the fit with the state at 11.23(8) MeV the red dashed line shows the fit without this state.

TABLE II: Comparison between experimental results and shell-model calculation. The first column shows the spin parity of the levels; the second column shows the experimentally determined excitation energies for the $^{18}$Ne levels observed in this work and in the literature. The references are reported in the table, the ($^*$) marks levels observed in the mirror nucleus $^{18}$O. The third column shows the experimental dimensionless reduced alpha width of the levels. The last four columns show the results of the shell-model calculation: Excitation energy, number of HO excitation quanta $\hbar\omega$, number of radial nodes in the relative motion wave function and the spectroscopic factors. Spectroscopic factors larger than 0.1 are highlighted in bold

| | Experiment | | Shell-model | | | |
|---|---|---|---|---|---|---|
| $J^\pi$ | $E_{exc}$ (MeV) | $\theta^2_a$ | $E_{exc}$ (MeV) | $\hbar\omega$ | n | SF |
| $0^+$ | $0^{[15,18]}$ | | 0 | 0 | 3 | **0.74** |
| $0^+$ | $3.6^{[15,19]}$ | | 3.4 | 2 | 4 | **0.62** |
| $0^+$ | $4.6^{[19-21]}$ | | 4.6 | 0 | 3 | **0.2** |
| $0^+$ | 9.8(3) | **1.4(6)** | 7.8, 11.8, 12.9, 13.3, 14.1 | 4 | 5 | <0.01, <0.01, <0.01, 0.02, **0.15** |
| $0^+$ | | | 7.9, 9.8, 10.4, 11.4, 11.6, 12.2 | 2 | 4 | <0.01, 0.01, 0.05, <0.01, <0.01, 0.07 |
| $1^-$ | $4.5^{[20,21]}$ | | 4.5 | 1 | 3 | **0.22** |
| $1^-$ | $6.2^{[20,21,23]}$ | | 6.7 | 1 | 3 | **0.12** |
| $1^-$ | $7.6^{[22,23]}$ | | 7.4 | 1 | 3 | **0.39** |
| $1^-$ | 9.13(2) | **0.22(2)** | 7.8, 9 | 1 | 3 | <0.01, 0.02 |
| $1^-$ | 9.61(3) | **0.52(5)** | 9.4, 9.8 | 3 | 4 | **0.47**, <0.01 |
| $1^-$ | 10.56(4) | **0.11(7)** | 10.4, 10.9 | 3 | 4 | **0.13**, <0.01 |
| $1^-$ | 11.74(5) | 0.09(1) | 11.5, 11.9, 12.1, 12.3, 12.8 13.6 | 1 | 3 | <=0.01 |
| $1^-$ | | | 12.4, 12.8, 13.2, 13.3 | 3 | 4 | <=0.01 |
| $1^-$ | 13.73(1) | **0.2(1)** | 13.2, 13.3 | 3 | 4 | <=0.01 |
| 1⁻ | 13.73(1) | **0.2(1)** | 11.8, 13.05 | 5 | 5 | N/A |
| 1⁻ | | | | | | |
| 2⁺ | 1.9[15,19] | | 2.0 | 0 | 2 | **0.64** |
| 2⁺ | 3.6[15, 19] | | 4.4 | 0 | 2 | **0.15** |
| 2⁺ | 5.1[19, 20] | | 5.1 | 2 | 3 | **0.62** |
| 2⁺ | (7.93(2)) | (0.12(5)) | 9, 9.47, 9.51 | 2 | 3 | <0.01 |
| 2⁺ | 9.21(3) | **0.21(2)** | 9.9, 11 | 0 | 2 | 0.08, **0.1** |
| 2⁺ | 10.8(1) | **0.55(3)** | 9.1, 10.7, 10.8 | 4 | 4 | <0.01 |
| 2⁺ | 13.4(2) | **0.45(8)** | 10.3, 10.5, 10.9, 10.94, 11.6, 11.7 | 2 | 3 | 0.03, 0.02, <0.01, 0.01, <0.01, <0.01 |
| 2⁺ | 13.4(2) | **0.45(8)** | 12.5, 12.8, 13.3, 13.8, 13.9, 14.4, 14.8 | 4 | 4 | <0.01 |
| 2⁺ | 16.26(2) | **0.2(1)** | 16.2 | 0 | 2 | 0.03 |
| 2⁺ | (16.9(2)) | (**0.3(2)**) | | | | |
| 3⁻ | 5.1[18, 20, 21] | | 5.1 | 1 | 2 | **0.48** |
| 3⁻ | 6.3[20, 23] | | 6.4 | 1 | 2 | 0.05 |
| 3⁻ | 8.29*[5] | | 7.8 | 1 | 2 | **0.13** |
| 3⁻ | 8.76(8) | **1.0(4)** | 8.95 | 3 | 3 | **0.14** |
| 3⁻ | 8.76(8) | **1.0(4)** | 8.98, 9.5, 9.9, 10.6 | 1 | 2 | 0.03, <0.01, 0.05, 0.06 |
| 3⁻ | 11.0(1) | **0.21(7)** | 10.7 | 3 | 3 | **0.33** |
| 3⁻ | | | 10.9, 11.1, 11.5 | 1 | 2 | <0.01 |
| 3⁻ | 12.13(4) | 0.07(3) | 12.1 | 3 | 3 | 0.09 |
| 3⁻ | 12.45(2) | 0.07(5) | 12.3, 12.9, 13.44 | 3 | 3 | 0.01, 0.03, 0.01 |
| 3⁻ | | | 12.6, 12.9 | 1 | 2 | <0.01 |
| 3⁻ | 12.7(2) | **0.7(2)** | 13.5, 13.7, 14.0 | 3 | 3 | 0.03, 0.02, <0.01 |
| 3⁻ | 12.7(2) | **0.7(2)** | 13.1, 13.5, 14, 14.2 | 1 | 2 | 0.03, 0.02, <0.01, <0.01 |
| 3⁻ | 14.8(2) | **1.0(2)** | 12.39 | 5 | 4 | N/A |
| 4⁺ | 3.4[15, 19] | | 3.5 | 0 | 1 | **0.43** |
| 4⁺ | (8.16$^{+0.05}_{-1.1}$) | (**0.8(3)**) | 7.9 | 2 | 2 | **0.54** |
| 4⁺ | | | 9.02 | 0 | 1 | **0.57** |
| 4⁺ | | | 10.2, 10.3, 11, 11.2, 11.8 | 2 | 2 | <0.01, 0.02, <0.01, 0.03, <0.01 |
| 4⁺ | | | 11.2, 11.6 | 4 | 3 | <0.01 |
| 4⁺ | | | 12.2, 12.5, 12.8, 12.9 | 2 | 2 | <0.01, <0.01, <0.01, 0.02 |
| 4⁺ | 13.3(3), 14.2(2) | **0.37(4), 0.14(10)** | 13.1, 14.2, 14.5, 14.7, 15.1, 15.4, 15.9, 16 | 4 | 3 | <0.01 |
| 5⁻ | 7.9*[9,18] | | 7.8 | 1 | 1 | **0.32** |
| 5⁻ | 8.1*[9,18] | | 8.4 | 1 | 1 | **0.36** |
| 5⁻ | 9.7*[18] | | 9.6 | 1 | 1 | <0.01 |
| 5⁻ | 11.31(4) | 0.03(2) | 11.8 | 3 | 2 | 0.02 |
| 5⁻ | | | 12.4, 12.7, 13.3 | 1 | 1 | <=0.01 |
| 5⁻ | 12.9(2) | 0.5(1) | 13.8, 14.1 | 1 | 1 | 0.05, 0.02 |
| 5⁻ | 13.79(8), 14.6(7) | 0.1(1), 0.3(2) | 13.7, 14.3, 14.7, 14.9 | 3 | 2 | 0.25, <0.01, 0.02, 0.04 |
| 5⁻ | 14.9(1) | 0.03(2) | 14.6, 14.7, 15.2 | 1 | 1 | <0.01, 0.02, 0.02 |
| 5⁻ | | | 13.9 | 5 | 3 | N/A |
| 5⁻ | | | 15.2, 15.3, 15.5 | 3 | 2 | 0.01, <0.01, 0.08 |
| 6⁺ | (11.23) | (0.04(3)) | 12.1 | 0 | 0 | 0.07 |
| 6⁺ | 11.8(2) | 0.23(7) | 12.5 | 0 | 0 | 0.44 |
| 6⁺ | 12.4(2) | 0.6(3) | 13.5 | 2 | 1 | <0.01 |
| 6⁺ | 12.4(2) | 0.6(3) | 13.7 | 0 | 0 | 0.01 |

**Discussion**

We studied the properties of the excited states in $^{18}$Ne in the excitation energy range from 8 to 17 MeV which were populated by resonant alpha elastic scattering on $^{14}$O. Given the reduced statistics of our data, this experiment is mostly sensitive to states with large alpha partial widths and small proton partial widths. As expected from isospin symmetry, there is a good correspondence of the $^{18}$Ne levels with those of the mirror nucleus $^{18}$O. The data show that alpha-cluster configuration plays a significant role.

The shell-model analysis allowed us to classify the observed states based on their particle-hole structure and relative alpha channel. The model reproduces the properties of the levels very well especially at excitation energies below 12 MeV. At higher excitation energies, the alpha strength in the model is generally fragmented into many states, while in the experiment we only observe a few broad states. This indicates that the structure of these broad states is collective and is aligned towards the corresponding alpha reaction channel.

Narrow near-threshold cluster-like resonances are known to appear at energies near the particle decay threshold. These resonances are very important especially in astrophysical settings. Two examples are the famous Hoyle state in $^{12}$C, located 287 keV above the alpha decay threshold, and the proton-emitting near-threshold resonance in $^{11}$B at 11.425 MeV, 197 keV above the one-proton emission threshold [39]. As discussed in [39] coupling to the reaction continuum is important in the formation of these states. The alpha-cluster states considered in this paper in $^{18}$Ne and $^{18}$O are quite far from the respective alpha decay thresholds (> 3 MeV and > 2 MeV). In this paper we suggest that superradiance can be responsible for the formation of the broad alpha cluster states observed in the experiment both in $^{18}$O and in $^{18}$Ne. Superradiance is a phenomenon related to the coupling with the continuum in open quantum systems. Therefore, in analogy with the near-threshold alignment mentioned above, superradiant states are aligned with certain alpha-cluster configurations as they become energetically available so that, for each configuration, we observe one broad superradiant state instead of many. The role of superradiance in the formation of alpha cluster states in $^{18}$O and $^{18}$Ne will be discussed in a separate paper [40]. We stress here that mirror nuclei are an ideal ground to study superradiance in nuclear systems. In fact, these nuclei have basically the same structure, but differ in their coupling with the continuum due to the different coulomb interaction in the decay channels. Examining different mirror pairs in the future is necessary to have a better understanding of this phenomenon.

Baba and Kimura suggested in ref. [25], that the excitation energy shift between mirror levels can be used to distinguish different geometrical configurations. The shell-model calculation presented in this paper is isospin symmetric, therefore, it cannot be used to study the excitation energy shift between mirror levels in $^{18}$Ne and $^{18}$O. However, shell-model particle-hole configurations can be related to AMD geometric configurations when comparing the calculations to the experimental data. In the following we try to connect the AMD cluster structures in $^{18}$O from ref. [41] with the shell-model configurations presented in this paper for $^{18}$O and $^{18}$Ne. Further study of $^{18}$Ne with AMD together with higher statistics experimental data are necessary in the future to better understand the relation between excitation energy shift in mirror states and alpha-cluster configurations. Baba and Kimura used AMD to investigate different geometrical cluster structures in $^{18}$O [41]. They found 5 different types of cluster states: $^{14}$C + $\alpha$, $^{14}$C + $\alpha$ higher nodal, two types of molecular states and a 4$\alpha$ linear-chain. One of the molecular states and the 4$\alpha$ linear-chain appear at energies above the limit of our data. They considered positive and negative parity states and compared the theoretical results with the $^{18}$O experimental data in the literature. The states with $^{14}$C + $\alpha$ configurations preferentially decay by alpha emission and have a large alpha spectroscopic factor, whereas the molecular states have small alpha spectroscopic factors and preferentially decay by $^{6}$He emission. Since there is a good correspondence between the levels with prominent alpha-cluster structure in $^{18}$O and $^{18}$Ne it is possible to compare the levels with the $^{14}$C+$\alpha$ configurations in [41] with the corresponding levels in $^{18}$Ne. The levels corresponding to the ground band in $^{18}$O (0$^{+}$ at 0 MeV, 2$^{+}$ at 1.98 MeV, 4$^{+}$ at 3.55 MeV) correspond to the lowest 0$\hbar\omega$ states in our shell-model calculation. The $^{14}$C+$\alpha$ positive parity states in ref. [41] (0$^{+}$ at 3.6 MeV, 2$^{+}$ at 5.25 MeV, 4$^{+}$ at 7.12 MeV in the $^{18}$O experimental results) are 2$\hbar\omega$ states in our calculation. Their $^{14}$C + $\alpha$ negative parity states (1$^{-}$ at 9.76 MeV, 3$^{-}$ at 12.98 and 14 MeV, 5$^{-}$ at 12.94 and 14.1 MeV in the $^{18}$O experimental data) have 3$\hbar\omega$ in our calculation. The higher nodal $^{14}$C + $\alpha$ states in ref. [41] (0$^{+}$ at 9.9 MeV, 2$^{+}$ at 12.21/12.8 MeV, 4$^{+}$ at 14.77 MeV in the $^{18}$O experimental data) correspond to 4$\hbar\omega$ in our calculation.

To further investigate our assignments, FIG. 8 shows the energy of the 2$\hbar\omega$, 3$\hbar\omega$ and 4$\hbar\omega$ states with large alpha spectroscopic factors as a function of the spin (J(J+1)). Although the 3$\hbar\omega$ and 4$\hbar\omega$ states show some splitting, the excitation energies show a linear dependence from J(J+1) indicating the possible existence of alpha-cluster rotational bands in $^{18}$Ne. We also note that these different configurations seem to have a very similar moment of inertia.

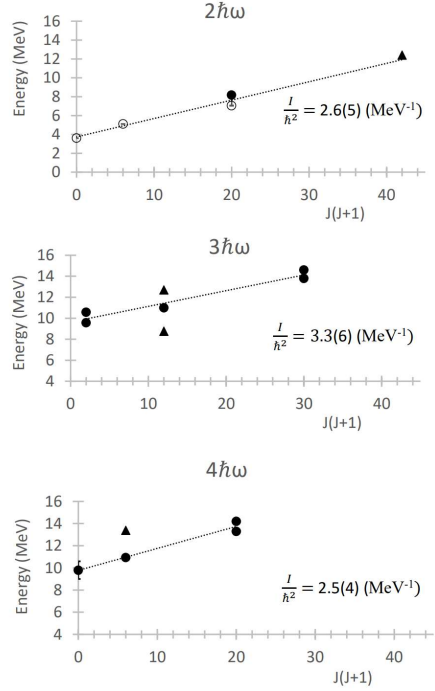

FIG. 8. Excitation energy as a function of the spin J(J+1) for the states with large alpha spectroscopic factor in the $2\hbar\omega$, $3\hbar\omega$ and $4\hbar\omega$ groups. Solid markers show our data whereas open circles show data from the literature. The solid triangles are states for which the assignment to the configuration is uncertain. The moments of inertia are reported in the figure.

**Conclusions:**

The resonant elastic scattering of $^{14}$O on $^{4}$He was used to measure the excitation function of $^{18}$Ne in inverse kinematics in the energy range from 8 to 17 MeV. A multi-channel R-matrix analysis of the data was performed starting from initial parameters derived from the known states in the mirror nucleus $^{18}$O [5]. Detailed spectroscopic information was obtained from the R-matrix fit of the $^{18}$Ne dataset (excitation energy of the states, spin and parity as well as total widths and partial alpha widths). A good overall correspondence between the levels in $^{18}$O and $^{18}$Ne was found, confirming the importance of isospin symmetry for alpha-cluster states in these N≠Z systems. The presence of several pairs of corresponding levels with large alpha dimensionless reduced width shows that alpha-clustering is strong in these nuclei. Comparing our experimental results with shell-model calculations allowed us to make significant progress in the microscopic understanding of clustering. We tried to categorize the states by their particle-hole structure and cluster channel.

The unmixed model reproduces the properties of the levels in $^{18}$Ne up to excitation energies of about 10-12 MeV. For each $J^\pi$ group, the model is indicating how the alpha strength is distributed among different particle-hole configurations and alpha-cluster decay channels. The results support the idea that the broad alpha-clustered states observed in the experiment account for most or the entire clustering strength in a particular channel, so that the strength of the corresponding particle-hole excitation levels is concentrated in one or few states. The superradiance mechanisms seems to play an important role in the generation of clustering in $^{18}$Ne and $^{18}$O. More experimental and theoretical efforts are necessary to further investigate superradiance's role in forming alpha-cluster states in N≠Z mirror pairs.

**Acknowledgments**

This work was supported by the United States Department of Energy under Grant No. DE-FG03-93ER40773 and No. DE-SC0009883 and by the UK Science and Technology Facilities Council (STFC) under Grant No. ST/P004199/1.

[1] J.A. Nolen Jr, J.P. Schiffer Annu. Rev. Nucl. Sci., **19**, 471 (1969)

[2] R.B. Wiringa et al. Phys. Rev. C **88**, 044333 (2013)

[3] P. Bączyk et al. Phys. Lett. B **778** 178-183 (2018)

[4] K.P. Artemov et al, Sov. J. Nucl. Phys. **52**, 406 (1990)

[5] M. L. Avila et al., Phys. Rev. C **90**, 024327 (2014)

[6] S. Pirrie et al. Phys. Rev. C **102**, 064315 (2020)

[7] S. Pirrie et al. Eur. Phys. J A **57**, 150 (2021)

[8] E.D. Johnson et al., Eur. Phys. J. A **42**, 135–139 (2009)

[9] W. von Oertzen et al., Eur. Phys. J. A **43**, 17–33 (2010)

[10] B. Yang et al. Phys. Rev. C **99**, 064315 (2019)

[11] A. Cunsolo et al. Phys. Rev. C **24**, 476 (1981)

[12] A. Osman, A.A. Farra Il Nuovo Cimento, Vol **103** N12, 1693 (1990)

[13] G.L Morgan et al. Nucl. Phys. A **148**, 480 (1970)

[14] G.L. Morgan et al. Phys. Lett. B **23**, 353 (1970)

[15] W.R. Falk et al. Nucl. Phys. A **157**, 241 (1970)

[16] S.H. Park et al. Phys. Rev. C **59**, 1182 (1999)

[17] Changbo Fu et al., Phys. Rev. C **77**, 064314 (2008)

[18] D. R. Tilley et al. Nucl. Phys. A **595**, 1-170 (1995)

[19] A.V. Nero, E.G. Adelberger, F.S. Dietrich, Phys. Rev. C **24**, 1864 (1981)

[20] K.I. Hahn et al., Phys. Rev. C **54**, 1999 (1996)

[21] R. J. Charity et al., Phys. Rev. C **99**, 044304 (2019)

[22] J.J. He et al., Eur. Phys. J. A **47**, 67 (2011)

[23] B. Harss et al. Phys. Rev. C **65** 035803 (2002)

[24] Konstantinos Kravvaris and Alexander Volya, Phys. Rev. C **100**, 034321 (2019)

[25] T. Baba and M. Kimura, Phys. Rev. C **99**, 021303 (R) (2019)

[26] M. Nakao et al., Phys. Rev. C **98**, 054318 (2018)

[27] E. Koshchiy et al., Nucl. Instr. and Methods in Phys. Res. A **957**, 163398 (2020)

[28] R.E. Tribble, R.H. Burch and C.A. Gagliardi, Nucl. Instr. And Methods in Phys. Res. A **285**, 441 (1989)

[29] Duda R.O. and Hart P.E. (1972). Use of the Hough transformation to detect lines and curves in pictures. Communications of the ACM, 15(1),11-15

[30] E.D. Johnson, The cluster structure of oxygen isotopes, Ph.D. thesis, Florida State University, Tallahassee, Florida, USA, 2008, proQuest UMI No. 3348500

[31] A.M. Lane and R.G. Thomas Rev. Mod. Phys. 30, 257 (1958)

[32] F. James MINUIT Function Minimization and Error Analysis: Reference Manual Version 94.1, Report Number: CERN-D-506 (1994)

[33] G.K. Tobin et al. Phys. Rev. C **89**, 034312 (2014)

[34] A. Volya et al. Phys. Rev. C **105**, 014614 (2022)

[35] R.S. Lubna et al. Phys. Rev. C **100**, 034308 (2019)

[36] R.S. Lubna et al. Physical Review Research 2, 043342 (2020)

[37] M. Avila et al., Phys. Rev. C **96**, 014322 (2017)

[38] A. Volya and V. Zelevinsky, J. Opt. B: Quantum Semiclass. Opt. **5**, S450 (2003)

[39] J. Okołowicz, M. Płoszajczak and W. Nazarewicz, Phys. Rev. Lett. 124, 042502 (2020)

[40] A. Volya et al., submitted to Nature Communications Physics, https://doi.org/10.21203/rs.3.rs-1926768/v1

[41] T. Baba and M. Kimura, Phys. Rev. C **100**, 064311 (2019)